# Optical surface waves over metallo-dielectric nanostructures: Sommerfeld integrals revisited

Bora Ung* and Yunlong Sheng

*Center for Optics, Photonics and Lasers, Laval University, Quebec, G1K 7P4, Canada*
(Dated: June 1, 2008)
*Corresponding author*: bora.ung.1@ulaval.ca

**Abstract:** The asymptotic closed-form solution to the fundamental diffraction problem of a linear horizontal Hertzian dipole radiating over the metallo-dielectric interface is provided. For observation points just above the interface, we confirm that the total surface near-field is the sum of two components: a long-range surface plasmon polariton and a short-range radiative cylindrical wave. The relative phases, amplitudes and damping functions of each component are quantitatively elucidated through simple analytic expressions for the entire range of propagation: near and asymptotic. Validation of the analytic solution is performed by comparing the predictions of a dipolar model with recently published data.

**1. Introduction**

Sparked by the discovery of enhanced optical light transmission through subwavelength holes pierced in a metal film [1], considerable interest has grown in the last decade for the guiding and manipulation of light in metallic nanostructures. A central role is attributed to electromagnetic surface waves (SW) that are launched by light diffraction on a structural nano-defect and subsequently propagated along the metallo-dielectric interface. While novel nano-devices operating on these surface-bound vectors are envisioned in many areas of science and engineering, fundamental-level research is still being performed in order to completely assess and control their intrinsic properties. In this respect, surface plasmon polaritons (SPP) have been identified early on to occupy a predominant role, especially in the asymptotic propagation regime. The SPP, which is a guided electromagnetic wave with fields evanescently extending on either side of the interface, represents a solution stemming from Maxwell's equations provided the metal substrate has a finite conductivity and the magnetic field is polarized transverse (TM) to the plane of incidence. An alternate type of SW denominated "composite diffracted evanescent wave" (CDEW), and characterized by a field damping scaling as $1/x$ with propagation distance $x$, has also been proposed [2]. The CDEW model is based on a number of approximations. First of all, it derives from a scalar-wave theory which ignores field polarization, and second, it assumes an opaque and infinitely thin screen as the boundary condition, thus precluding any likely involvement of the SPP mode. A key conceptual difference between the CDEW and the classical SPP model is that the former allows several propagating surface modes to be generated by diffraction while the latter only admits the SPP. Both models were applied to interpret recent experimental investigations of the near field with relative success [3-5]: the CDEW is ostensibly better suited in the immediate vicinity of the source whereas the SPP is most accurate further away from it. The indication that each theory presents complementing pictures of the same phenomenon was made apparent by Lalanne and Hugonin [6] when they suggested that the composite diffracted surface wave essentially consists of two components: a SPP and a radiative "creeping wave" characterized by a damping scaling with $1/x^{1/2}$. In their demonstration, the authors used a rigorous Green's function formalism to describe the field radiated by a line source over a metallo-dielectric half-space. However, no closed-form solution for the creeping wave was provided. New experimental results [7] taken from direct measurements of the surface near field have lent further credence to the thesis of a SW dual composition.

In this theoretical contribution, we demonstrate that the scattering process of incident TM-polarized light by a one-dimensional subwavelength nano-defect on an otherwise flat metal surface is generically modeled through the fundamental diffraction problem of a horizontal Hertzian dipole radiating over the metallo-dielectric interface. Starting directly from Maxwell's equations and enforcing the proper boundary conditions, we write down the exact Sommerfeld-type integral for the electric field perpendicular to the interface. The integral is then rigorously solved via the modified method of steepest descents to obtain the asymptotic closed-form solution. We explicitly show that the total diffracted field along the surface is composed of two distinct contributions as predicted in [6]: a short-range radiative wave and a long-range SPP. The amplitudes, phases and damping functions of both components are quantitatively revealed through simple analytic expressions. Finally, we propose a double-dipole model to characterize the near-field interactions between two nano-objects and show the excellent agreement between our model's predictions and the experiments, thus validating the analytical solution in the same stretch.

## 2. Derivation of the closed-form solution

It was demonstrated that the scattering of incident TM-polarized light on a subwavelength metallic slit induces oscillating electric charges around the sharp slit edges that emulate a horizontal electric dipole [8,9]. It can be argued that this distinctive feature equally applies to the subwavelength groove if it is considered as a partially filled slit. Hence, the physical problem of light scattering on a nano-slit, or nano-groove, can be represented by the basic diffraction problem of an infinitely small horizontal electric dipole radiating over the interface. That assumption is corroborated by other analytical and experimental investigations [10,11]. From our calculations, this approximation is fairly accurate if the width $w$ of the given nano-object is smaller than a half-wavelength, $w<\lambda/2$, and if observation points are located at a distance $x$ from the source larger than $x>\lambda/\pi$. The remaining discussion in this Section follows the contemporary mathematical treatment derived by R. E. Collin for the distinct case of a 3D vertical point dipole radiating over the earth's surface [12]; which was the original diffraction problem famously addressed by Arnold Sommerfeld one century ago.

*2.1 Expression of the Sommerfeld integral for a linear horizontal Hertzian dipole*

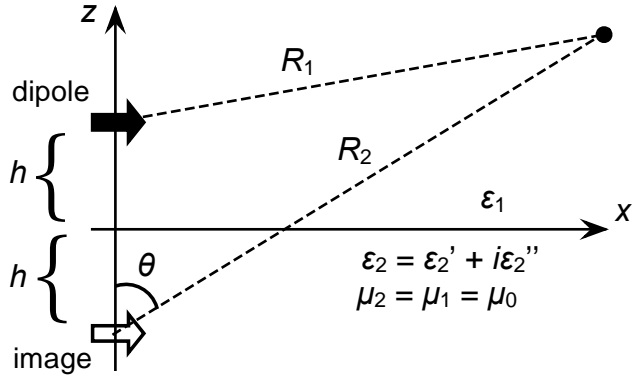

Fig. 1. Radiating horizontal electric dipole located in region 1 (air).

We consider the 2D linear electric dipole oriented along the $x$-axis and located in the half-space $z \geq 0$ at a height $h$ from the interface separating two semi-infinite nonmagnetic and isotropic dielectrics (Fig. 1). The infinitesimal line dipole is described by its surface current density $J_x=\delta(x)\cdot\delta(z-h)$ normalized to unit electric moment. We will not cover here the details of the straightforward though lengthy derivation of Eq. (1), which can be found in textbooks [13], but rather briefly summarize the procedures therein. After enforcing the boundary conditions at the interface for the $E_x$, $E_z$ and $H_y$ fields (TM-polarization) in Maxwell's curl equations then taking the inverse Fourier transform of the fields and choosing the appropriate Green's function, one derives the Sommerfeld-type integral for the normal electric field in region 1 ($z \geq 0$):

$$E_{1z} = \frac{\omega\mu_0}{2\pi k_0^2} \int_{-\infty}^{+\infty} \left[ \frac{\gamma_2 e^{i\gamma_1 h}}{\varepsilon_2\gamma_1 + \varepsilon_1\gamma_2} + \frac{\sin(\gamma_1 h)}{i\varepsilon_1} \right] e^{i\gamma_1 z} k e^{ikx} dk \qquad (1)$$

where $k$ is the lateral component of the total wavevector which is related to the normal component $\gamma$ in each region "$i$" via $\gamma_i = \sqrt{k_i^2 - k^2}$ with $k_i = k_0\sqrt{\varepsilon_i}$ and $k_0 = 2\pi/\lambda$. A time dependence in $e^{-i\omega t}$ is assumed in the present analysis and suppressed throughout.

*2.2 Asymptotic solution to the integral via the modified method of steepest descents*

The integral in Eq. (1) is separated in two parts, $E_{1z}=K(I_1 + I_2)$, with $K=\omega\mu_0/2\pi$, and $I_1$ and $I_2$ respectively corresponding to the first and second term inside the brackets. We first consider the integral $I_1$ and impose $\gamma_1$ and $\gamma_2$ to have positive imaginary parts in order for the field to be bounded at infinity. In the corresponding first quadrant of the complex $k$-plane, we discard the branch cut running from the point $k=k_2$ since its contribution – characterized by an attenuation factor $\exp(-\text{Im}\{k_2\}\cdot x)$ that drops to nearly zero within a quarter-wavelength distance $x$ – is negligible for the optical frequencies of interest compared to that of the branch cut running from $k=k_1$. There is also a pole singularity in the denominator of $I_1$ located at $\pm k_p = \pm k_0\sqrt{\varepsilon_p}$ where $\varepsilon_p = \varepsilon_1\varepsilon_2/(\varepsilon_1 + \varepsilon_2)$ defines the SPP's "effective permittivity". The proper pole ($+k_p$) lies very close to the branch point $k_1$ and we will see later on that this aspect entails particular considerations. We make the simplification $\varepsilon_1=1$ such that $k_1=k_0$ and perform the successive transformations $k = k_0 \sin\alpha$, $x = R_2 \sin\theta$ and $(z + h) = R_2 \cos\theta$, where $R_2$ is the distance from the mirror image of the dipole in region 2 to the observation point. These procedures enable one to express $I_1$ as the integral over an angular spectrum of plane waves:

$$I_1 = \int_C \frac{\sin\alpha \cdot \cos\alpha \cdot \sqrt{\varepsilon_2 - \sin^2\alpha} \cdot e^{ik_0 R_2 \cos(\alpha-\theta)}}{\left(\varepsilon_2 \cos\alpha + \sqrt{\varepsilon_2 - \sin^2\alpha}\right)} \cdot d\alpha \qquad (2)$$

The complex variable $\alpha = \sigma + i\eta$ defines the angle between the direction of propagation and the $x$-axis. The saddle-point is found by setting the derivative of the argument in exponential equal to zero, thus yielding $\alpha = \theta$ where $\theta$ is real. The location of the pole in the $\alpha$-plane is given by $k_p = k_0 \sin\alpha_p$. As a numerical example, at the excitation wavelength $\lambda=852nm$ and $\varepsilon_2=-33.22+1.17i$ (silver metal) we have $\alpha_p = \pi - \text{Arcsin}(k_p/k_0)=1.574-i0.175$ where $\text{Arcsin}(v) = \int_0^v dy/\sqrt{1-y^2}$. As shown on Fig. 2, the original integration contour $C$ is then deformed into the steepest-descent contour ($SDC$) whose path, $\cos(\sigma - \theta)\cosh\eta = 1$, is shifted to pass through the saddle-point at $\theta = \pi/2$ where the highest accuracy is assigned to observation points along the surface. Since the pole $\alpha_p$ has been crossed by the path in the process, and is positioned below the *SDC* and very close to the saddle-point (see Fig. 2), the pole is captured and its contribution must be accounted for with a residue [12,14].

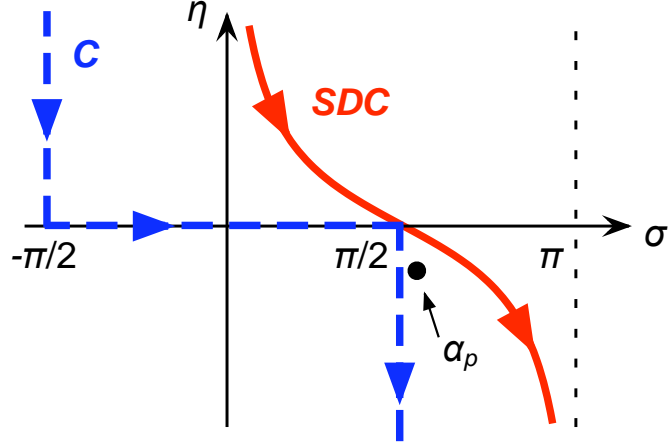

Fig. 2. Integration contours and location of the single pole in the α-plane.

To carry out the steepest-descent integration in Eq. (2), we first perform the change of variable $\tau = 2e^{i\pi/4}\sin((\alpha-\theta)/2)$, which yields the relations $d\alpha/d\tau = e^{-i\pi/4}/\cos((\alpha-\theta)/2) = -2i/\sqrt{\tau^2-4i}$, such that the integral $I_1$ is rewritten:

$$I_1 = e^{ik_0 R_2} \int_{-\infty}^{+\infty} f(\tau) \cdot \exp\left(-k_0 R_2 \frac{\tau^2}{2}\right) \cdot d\tau \qquad (3)$$

where $f(\tau) = g(\tau)/h(\tau)$ and:

$$g(\tau) = \sin\alpha \cdot \cos\alpha \cdot \sqrt{\varepsilon_2 - \sin^2\alpha} \cdot \frac{d\alpha}{d\tau} \qquad (4)$$

$$h(\tau) = \varepsilon_2 \cos\alpha + \sqrt{\varepsilon_2 - \sin^2\alpha} \qquad (5)$$

Because a pole singularity lies very close to the saddle-point $\theta = \pi/2$, the classical steepest-descent method cannot be applied directly. To circumvent this problem, the procedure that we will employ here is a modified steepest-descent technique described in [12]. In this approach we separate the pole and analytical contributions, respectively $f_P(\tau)$ and $f_A(\tau)$, by subtracting out the pole term from the main integrand:

$$f(\tau) = \frac{A}{\tau - \tau_p} + \frac{f(\tau) \cdot (\tau - \tau_p) - A}{\tau - \tau_p} \qquad (6)$$

$$= f_P(\tau) + f_A(\tau)$$

where $f_P(\tau)$ is the pole term with residue constant *A*. The remaining $f_A(\tau)$ is holomorphic up to the next singular point at $\tau^2 = 4i$ which arises in the term $d\alpha/d\tau$ present in $g(\tau)$. The function $f_A(\tau)$ is expanded in a Taylor series around the saddle-point $\tau = 0$ (i.e. $\alpha = \theta$) where we keep only the first term and neglect the remaining higher even-order correction terms to obtain $f_A(\tau) \approx f_A(0) = f(0) + A/\tau_p$. By evaluating the remainder's upper bound, we estimated the typical maximum relative error associated with the truncation of the Taylor series at the first term to be below 12%. With the help of the relations $\sqrt{\varepsilon_2 - \sin^2 \alpha_p} = -\varepsilon_2 \cos \alpha_p$ and $\cos^2 \alpha_p = (1+\varepsilon_2)^{-1}$, we find the value of the constant $A$:

$$A = \frac{g(\tau_p)}{\left.\frac{\partial}{\partial \tau} h(\tau)\right|_{\tau_p}} = \frac{g(\tau_p)}{\left.\frac{\partial}{\partial \alpha} h(\alpha) \cdot \frac{\partial \alpha}{\partial \tau}\right|_{\tau_p}} = \frac{\varepsilon_p^2}{\varepsilon_2 - 1} \tag{7}$$

We separate the integral $I_1$ in two parts, $I_1 = I_P + I_A$, where $I_P$ and $I_A$ respectively denote the pole and analytical contributions from $f_P(\tau)$ and $f_A(\tau)$. We first consider the analytical part:

$$I_A = e^{ik_0 R_2} \int_{-\infty}^{+\infty} f_A(0) \cdot \exp\left(-k_0 R_2 \frac{\tau^2}{2}\right) \cdot d\tau \tag{8}$$

Since the function $f_A(0)$ has no dependence in $\tau$, it can be removed from the integrand. Upon evaluating the remaining kernel, $\int_{-\infty}^{+\infty} \exp(-k_0 R_2 \tau^2/2) \cdot d\tau = \sqrt{2\pi/k_0 R_2}$, we obtain:

$$I_A = \left[\frac{e^{-i\pi/4} \sin\theta \cos\theta \cdot \sqrt{\varepsilon_2 - \sin^2 \theta}}{\left(\varepsilon_2 \cos\theta + \sqrt{\varepsilon_2 - \sin^2 \theta}\right)} + \frac{Ae^{-i\pi/4}}{2 \cdot \sin((\alpha_p - \theta)/2)}\right] \sqrt{\frac{2\pi}{k_0 R_2}} e^{ik_0 R_2} \tag{9}$$

Next we compute the pole contribution:

$$I_P = e^{ik_0 R_2} \int_{-\infty}^{+\infty} f_P(\tau) \cdot \exp\left(-k_0 R_2 \frac{\tau^2}{2}\right) \cdot d\tau = A e^{ik_0 R_2} P(\tau) \tag{10}$$

where $P(\tau) = \int_{-\infty}^{+\infty} d\tau \cdot e^{-\chi^2 \tau^2}/(\tau - \tau_p)$ and $\chi = \sqrt{k_0 R_2/2}$. Special care must be taken to evaluate $P(\tau)$ since it can be defined by different functions depending on whether the pole is located in the lower-half (Im{$\tau_p$}<0) or upper-half (Im{$\tau_p$}>0) $\tau$-plane. We can find a solution valid over the entire $\tau$-plane by first evaluating $P(\tau)$ for Im{$\tau_p$}>0 and then assuming Im{$\tau_p$}<0 in the resulting integral expression. In which case we obtain $P(\tau) = i\pi e^{-\chi^2 \tau_p^2}\left[2 - \frac{2}{\sqrt{\pi}} \int_{-t}^{\infty} e^{-t^2} dt\right]$ with $t = -i\chi\tau_p$, such that:

$$I_P = 2\pi i \cdot A \cdot U(R_2, \theta) \cdot \exp\left[ik_0 R_2 \cos(\alpha_p - \theta)\right] \tag{11}$$

where: $$U(R_2, \theta) = 1 - \frac{1}{2} erfc\left[(1+i)\sqrt{\frac{k_0 R_2}{2}(\cos(\alpha_p - \theta) - 1)}\right] \tag{12}$$

and $erfc(v) = \frac{2}{\sqrt{\pi}} \int_v^\infty e^{-y^2} dy$ denotes the complementary error function. For the complete discussion regarding the evaluation of the pole integral $P(\tau)$, we refer the reader to the Appendix of [12]. The integrand in the next integral, $I_2$, is an entire function; therefore the standard method of steepest descents can be applied to it. With the identity $\sin(\gamma_1 h) = (e^{i\gamma_1 h} - e^{-i\gamma_1 h})/2i$, this last integral is separated in two parts, $I_I$ and $I_D$, respectively designating the image and direct dipole contributions. Each part is then solved following the same procedure previously described for $I_A$ with the exception that the transformations $x = R_1 \sin\theta$ and $(z - h) = R_1 \cos\theta$ are substituted in the single case of $I_D$. The solutions for $I_I$ and $I_D$ then represent free-space cylindrical waves emanating from their respective image and direct dipole origins:

$$I_I = -\frac{\sin\theta \cos\theta \cdot e^{-i\pi/4}}{2}\sqrt{\frac{2\pi}{k_0 R_2}}\exp(ik_0 R_2) \qquad (13)$$

$$I_D = +\frac{\sin\theta \cos\theta \cdot e^{-i\pi/4}}{2}\sqrt{\frac{2\pi}{k_0 R_1}}\exp(ik_0 R_1) \qquad (14)$$

The complete air-side normal $E$-field is written $E_{1z} = K(I_P + I_A + I_I + I_D)$. Upon adding the terms $I_A$ and $I_I$ together and performing some algebraic manipulations, one obtains the following general closed-form expression:

$$E_{1z} = \frac{\omega\mu_0}{4\pi}\sin\theta \cos\theta \cdot e^{-i\pi/4}\sqrt{\frac{2\pi}{k_0 R_1}}e^{ik_0 R_1} \qquad \rightarrow \text{Direct dipole field} \qquad (15)$$

$$+\frac{\omega\mu_0}{4\pi}\sin\theta \cos\theta \cdot r_{TM} \cdot e^{-i\pi/4}\sqrt{\frac{2\pi}{k_0 R_2}}e^{ik_0 R_2} \qquad \rightarrow \text{Reflected field} \qquad (16)$$

$$+\frac{\omega\mu_0}{4\pi} A \cdot e^{i\pi/4}\sqrt{\frac{4\pi}{k_0 R_2\left(\cos(\alpha_p - \theta) - 1\right)}}e^{ik_0 R_2} \qquad \rightarrow \text{Boundary wave field} \qquad (17)$$

$$+\omega\mu_0 A \cdot e^{i\pi/2} \cdot U(R_2,\theta) \cdot \exp\left[ik_0 R_2 \cos(\alpha_p - \theta)\right] \qquad \rightarrow \text{SPP field} \qquad (18)$$

Equations (15)-(16) define the geometrical-optics field while Eqs. (17)-(18) describe the diffracted field. The term $r_{TM} = \left(\sqrt{\varepsilon_2 - \sin^2\theta} - \varepsilon_2 \cos\theta\right)\big/\left(\sqrt{\varepsilon_2 - \sin^2\theta} + \varepsilon_2 \cos\theta\right)$ present in Eq. (16), is the Fresnel reflection coefficient for TM-polarized light incident on the air-metal interface. The function $U(R_2,\theta)$, found in Eq. (18) and determined in Eq. (12), defines a complex envelope multiplying the SPP phasor.

### 3. Analysis of the surface near field

*3.1 Reduction of the general solution for observation points along the surface*

For the critical case of observation points just above and parallel the interface at $\theta = \pi/2$, we get $x = R_2 \sin\theta = R_2$ and the geometrical-optics field components [Eqs. (15)-(16)] vanish while the diffracted field components [Eqs. (17)-(18)] remain. The total near field along the interface then describes a composite surface wave (SW) created by two co-propagating vectors: a surface plasmon polariton (SPP) evanescent wave and a "boundary wave" (BW) having essentially a free-space cylindrical nature. The boundary wave lies in the geometrical shadow so as to compensate for the discontinuity in the geometrical-optics field across the planar interface [15]. We refrain from using the denomination "creeping wave", which was previously chosen in [6], to identify this radiative wave since it generally refers to a surface mode propagating in the geometrical shadow around convex surfaces. The contributions of the SPP and BW respectively correspond to the first and second terms in Eq. (19). In that expression, the "mismatch parameter" $\Delta k = (k_p - k_0)$ defines the difference between the evanescent and free-space wavevectors. We normalize the total surface wave field, which we denote $E_{sw}$, by taking out the factor $\omega\mu_0 A e^{i\pi}$ from Eqs. (17)-(18) and moving the $\sqrt{x}$ root from the denominator to the numerator in order to eliminate any sign ambiguity when $x < 0$:

$$E_{sw}^+(x) = U(x) \cdot \exp[i(k_p x - \pi/2)] + \frac{1}{\sqrt{4\pi \cdot \Delta k}} \frac{\sqrt{x}}{x} \exp[i(k_0 x - 3\pi/4)] \qquad (19)$$

$$= E_{spp}^+(x) + E_{bw}^+(x)$$

The + superscript in Eq. (19) indicates that the SW originates on the right of the source dipole and propagates in the +x direction. An identical SW is excited in the –x direction whose field is the complex conjugate: $E_{sw}^-(x) = \overline{E}_{sw}^+(x)$. We note that the SPP and BW are initially phase-shifted by $\pi/4$. One may also notice that the expression of the BW involves the asymptotic form of the first-order Hankel function of the first kind, $H_1^{(1)}(k_0 x) \approx \sqrt{2/(\pi k_0 x)} \exp[i(k_0 x - 3\pi/4)]$, which is accurate for $k_0 x > 2$. This outcome is not incidental because the steepest-descent method – which is asymptotically exact – was used to obtain the solution. Thus in principle, the BW could be written in "exact" form with the Hankel function, $E_{bw}^+(x) = \sqrt{k_0/(8 \cdot \Delta k)} \cdot H_1^{(1)}(k_0 x)$, which highlights the cylindrical character of the BW. From the preceding remarks we expect Eq. (19) to be likewise accurate for $k_0 x > 2$.

## 3.2 Fresnel diffraction effects and the asymptotic propagation regime

The inspection of Eq. (19) reveals that the SPP excited through diffraction at the interface is not pure but rather modulated by a slowly oscillating envelope $U(x)$ owing to Fresnel diffraction effects [Eq. (12) evaluated at $\theta = \pi/2$]:

$$U(x) = 1 - \frac{1}{2} erfc\left[(1+i)\sqrt{\frac{\Delta k \cdot x}{2}}\right] \qquad (20)$$

The first term on the right-hand side of Eq. (20) defines the classical SPP mode whereas the second *erfc* term describes the fringe pattern generated by the process of coupling incident homogeneous light into the evanescent mode via scattering. As a note, Eq. (20) can equivalently be expressed using the conventional complex Fresnel integral, $F(v) = \int_0^v \exp(i\pi y^2/2) \cdot dy$, instead of the *erfc*: $U(x) = \frac{1}{2} + \frac{1}{2}(1-i) \cdot F(i\sqrt{\Delta k \cdot 2x/\pi})$. As shown on Fig. 3, the modulation function $Re\{U(x)\}$ evolves from $U(0)=0.5$ at the origin before reaching the first peak at some distance $x_1$ and then oscillates onward with a large distinctive period $\Delta L = 2\pi/|\Delta k|$ to lower amplitudes before asymptotically growing towards ever higher amplitudes due to the positive imaginary component inside the argument of *erfc*.

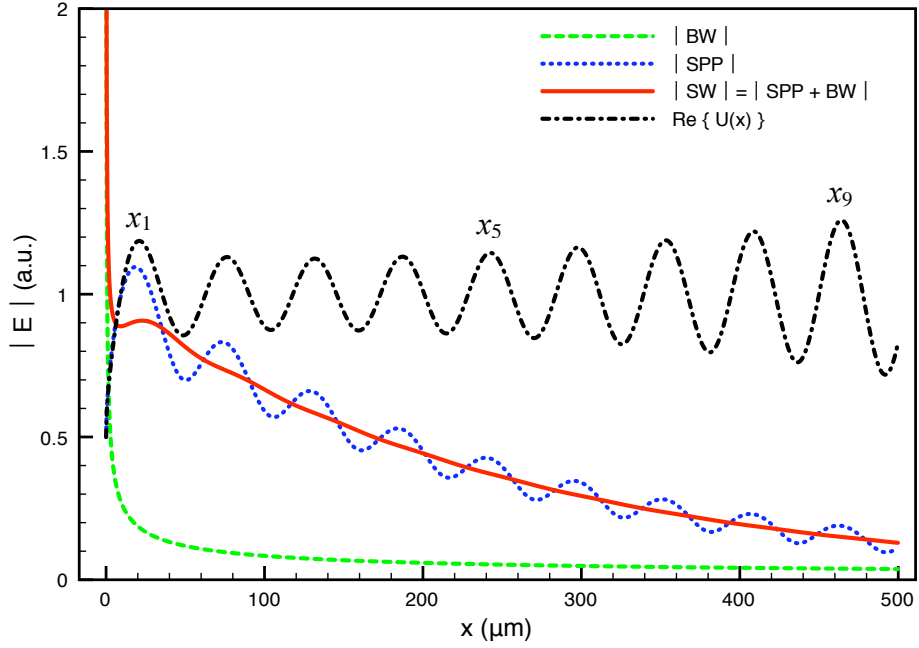

Fig. 3. Amplitude moduli of the total field (solid line), SPP field (dotted line), boundary wave field (dashed line) and the envelope function $U(x)$ (dashed-dotted line) for $\varepsilon_1=1$, $\varepsilon_2=-33.22+i1.17$ and $\lambda=852nm$. The successive peaks of $U(x)$ are numbered $x_1, x_2, ..., x_m$.

The location of the *m*-th peak from the origin is predicted by $x_m = \pi(2m-5/4)/|\Delta k|$ with $m=1,2,3...$; which yields $x_1=20.8\mu m$, $x_5=242.2\mu m$ and $x_9=463.6\mu m$ for the example of Fig. 3. The complex modulated SPP field is superposed on the monotonously decaying BW. As shown on the same figure, the sum of the two co-propagating surface waves produces a much weaker modulation, also with period $\Delta L$, of the total amplitude. Clearly, this beating behavior – more discernible at higher optical frequencies – is related to the wavevector mismatch between the evanescent and homogeneous modes, for when $\Delta k \to 0$ in the PEC limit (perfect electric conductor) all fringing effects disappear. The most salient feature is the first peak at $x_1 = 3\pi/(4|\Delta k|)$ which may represent the only detectable Fresnel modulation of the total field in practice due to the typically low visibility of the following fringes. To our best knowledge, this phenomenon of field pattern created by the interference of the two co-propagating surface waves – one of which is the evanescent SPP and the other a radiative wave – both originating from the same source, has not yet been reported in the literature.

*3.3 Characterization of the field at the origin and in the near-zone propagation regime*

The first point away from the origin where the real part of the oscillating BW field takes a zero value is approximately located at $x=\lambda/8$, yielding: $E_{bw}^+(\lambda/8) = \sqrt{k_0/(\pi^2 \Delta k)} \cdot e^{-i\pi/2}$. For $x \geq \lambda/8$ the real part is well-behaved; while for $x < \lambda/8$ it rapidly diverges towards negative values to become singular at the limit $x \to 0^+$. By contrast, the SPP mode is well defined at $x=0$: $E_{spp}^+(0) = U(0) \cdot e^{-i\pi/2}$. These previous remarks suggest that the SW field at the origin can be reasonably approximated by: $E_{sw}^+(0) \approx E_{spp}^+(0) + E_{bw}^+(\lambda/8) = \left[\frac{1}{2} + \sqrt{k_0/(\pi^2 \Delta k)}\right] \cdot e^{-i\pi/2}$.

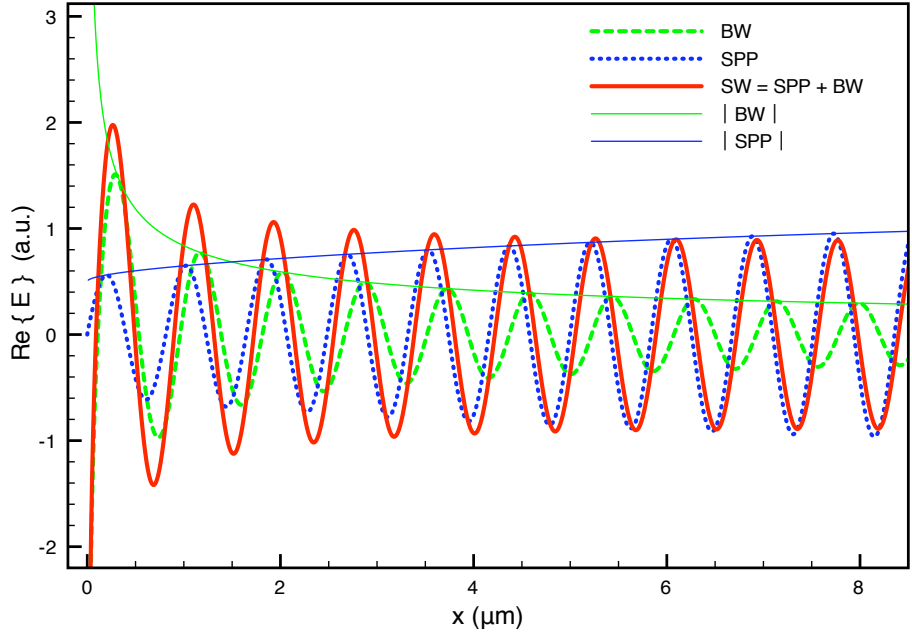

Fig. 4. Real components of the total field (solid line), boundary wave field
(dashed line), SPP field (dotted line) and their moduli for $\lambda/8 \leq x \leq 10\lambda$

The BW's amplitude scales as $1/(4\pi|\Delta k| \cdot x)^{1/2}$ in accordance with the $1/x^{1/2}$ damping predicted in [6]; while the SPP modulus follows $\text{Re}\{U(x)\} \cdot \exp(-\text{Im}\{k_p\}x)$. The "critical distance" where the moduli of the SPP and BW coincide is approximately located at $x_c = 4\pi/(67|\Delta k|)$. For the conditions $\lambda = 852 nm$ and $\varepsilon_2 = -33.22 + 1.17i$, the predicted value is $x_c = 1.64 \mu m$. For propagation lengths below the critical distance ($0 < x < x_c$) the BW provides the dominant contribution whereas the long-range SPP is the main vector thereafter (Fig. 4). The $1/e$ intensity decay lengths of the SPP and BW are respectively $L_{spp} = 1/(2 \cdot \text{Im}\{k_p\})$ and $L_{bw} = e/(4\pi|\Delta k|)$. For the same previous conditions, the BW's characteristic decay length $L_{bw} = 1.9 \mu m$ is much shorter compared to the SPP's $L_{spp} = 122 \mu m$. As a consequence, the effective value ($k_{sw} = 2\pi/\lambda_{sw}$) of the composite SW's wavevector (or effective index: $n_{sw} = k_{sw}/k_0$) is significantly affected by the rapid decay of the BW inside the near-zone regime ($0 < x < 2L_{bw}$) as shown in Fig. 4. The behavior of the near field depicted by the analytical solution in the source origin's vicinity, brings a clear and cohesive physical explanation to the reported accounts of transient phenomena in the first few wavelengths of propagation [3,4,7,10].

## 4. Dipolar model of the near-field interactions between nano-objects

As indicated earlier, when light is incident on a nano-object (nano-slit or nano-groove) with dimensions smaller than the wavelength of light, the scattered far-field radiation from the nano-object can be considered as originating from an infinitesimal horizontal electric dipole (HED) located on the structure. This assumption has been rigorously validated in Green's tensor numerical calculations by Lévêque et al. [10]. Hence in principle, many diffraction problems involving nanostructured surfaces can be modeled by replacing the individual nano-objects with point-sized dipoles. Indeed we demonstrate in this Section that recently reported experiments using groove-slit and double-slit configurations are accurately modeled by placing radiating linear HEDs at the corresponding sites of the surface nano-defects.

*4.1 Groove-slit transmission*

We first consider the groove-slit experiment performed by Gay *et al*. [3] and theoretically investigated by Lalanne and Hugonin [6]. The main interactions in this setup are described by the simple model illustrated in the diagram below (Fig. 5) where the groove-slit center-to-center separation distance is controlled by the variable *d*.

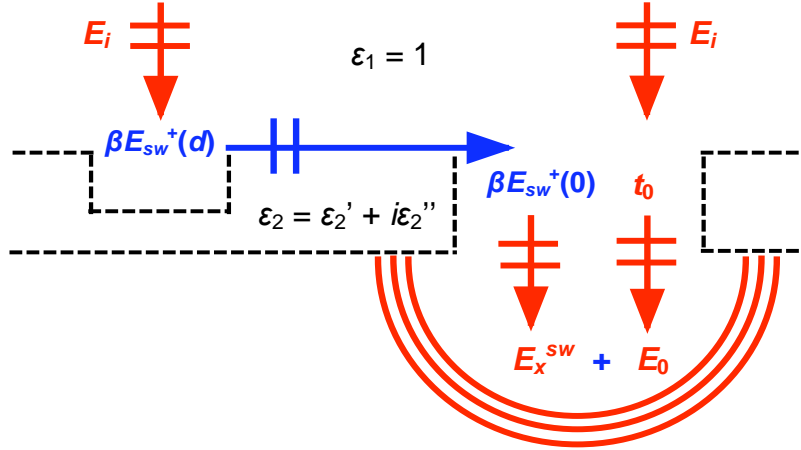

Fig. 5. Schematic of the near-field interactions in the groove-slit setup

An incident *x*-polarized plane wave of amplitude $E_i$ is scattered on both 100*nm*-wide groove and slit as HEDs. The fraction of incident light that is coupled by the groove into a *z*-polarized +*x*-directed SW is determined by the multiplication factor $\beta E_{sw}^+(d)$ where $\beta$ is the amplitude-coupling coefficient and $E_{sw}^+(d)$ describes the field propagation of the excited SW. The diffraction-launched SW then impinges on the adjacent nano-object (i.e. slit) where it is partially reflected and re-radiated. The re-radiated *x*-polarized component $E_x^{sw}$ that is coupled into the slit is defined by the factor $\beta E_{sw}^+(0)$, where backconversion reciprocity (SW-to-radiation and vice-versa) is assumed for the coupling coefficient $\beta$ and the factor $E_{sw}^+(0)$ accounts for the generation of a new SW along the slit's left-wall. Indeed, as demonstrated in [9] a new SW originates from the slit's top-left corner and whose initial value, $E_{sw}^+(0) \approx \left[\frac{1}{2} + \sqrt{k_0/(\pi^2 \Delta k)}\right] \cdot e^{-i\pi/2}$, introduces a $\pi/2$ phase-shift in magnitude. This intrinsic $\pi/2$ phase-shift between the SW generated at the groove and the directly incident light is consistent with earlier experimental [3,4] and theoretical studies [10]. The *x*-polarized light normally incident on the slit similarly generates two SWs: one in the forward and another in the backward *x*-direction. As described in [9], the SW launched from the groove does not directly interfere with the normally incident light since they have orthogonal polarizations; instead it is the re-radiated *x*-polarized component $E_x^{sw}$ that interferes with the transmitted component $E_0$ in the slit. The complete process of interference at the slit, as described by this simple model, is expressed by $I = |E_x^{sw} + E_0|^2$ where $E_x^{sw} = \beta E_{sw}^+(0) \cdot \beta E_i \cdot E_{sw}^+(d)$ and $E_0 = t_0 E_i$ respectively denote the *x*-polarized transmitted components relating to the left incident SW and the normal incident plane wave. The total transmitted intensity *I* is a function of groove-slit distance *d*, and is normalized with that without adjacent groove ($I_0 = |E_0|^2$) such as performed in [3,16]. Figure 6 shows the strong correlation between the experimental data [3], the fully-vectorial numerical calculations [16] and the predictions of the analytical dipolar model with parameters $\lambda=852nm$, $\varepsilon_1 = 1$, and $\varepsilon_2=-33.22+1.17i$. The approximate values of the scattering and modal transmission coefficients, respectively $\beta=0.357$ and $t_0=1.40$, are both calculated using the semi-analytical SPP generation model of Lalanne *et al.* [17].

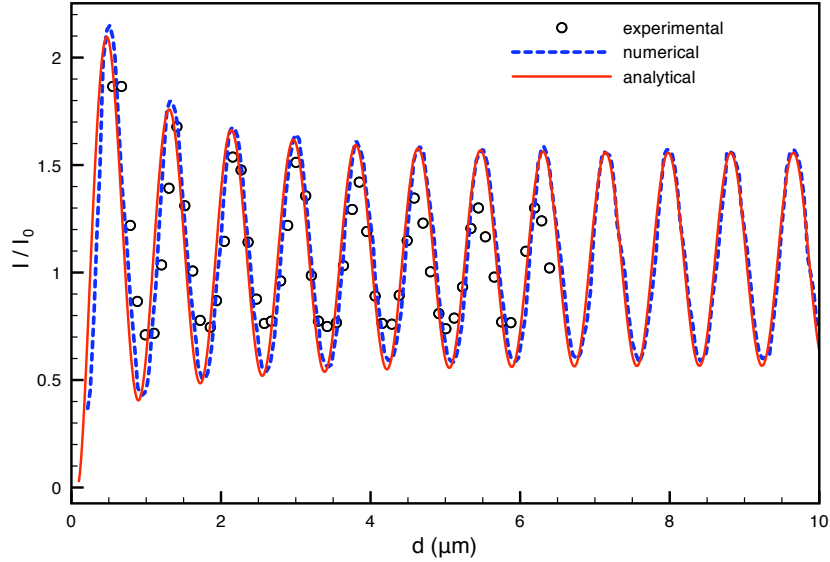

Fig. 6. Normalized slit transmission as a function of groove-slit distance *d*. Comparison between experimental (circles) [3], numerical (dashed line) [16], and the analytical (solid line) results with $\lambda=852nm$ and $\varepsilon_2 = -33.22+1.17i$.

## 4.2 Double-slit near field

The field just above the surface between two subwavelength slits milled in a gold film, separated by $2d=10.44\mu m$ and illuminated with $\lambda=974.3nm$ normal incident TM-polarized light, was recently measured through a scanning near-field optical microscope (SNOM) by Aigouy and co-workers [7]. Neglecting the tangential component of the *E*-field at the interface, the patterns recorded from fluorescence emission are expected to scale with $|E_z|^4$. Modeling once more each nano-object (i.e. the slits) as a linear HED, the resulting near-field intensity pattern arises from the interference of two counter-propagating SWs, $E_z(x) = E_{sw}^+(x+d) + E_{sw}^-(x-d)$, where the origin of the *x*-axis is located at the half slit-to-slit separation distance. In Fig. 7 the near field between the slits, $a|E_z(x)|^4 + b$, is plotted in the vertical axis, where $b=0.33$ is the background illumination offset taken from the original data and $a=0.031$ is the best-fit gain factor. The decay, phase and pattern periodicity predicted by the analytical model, with $\varepsilon_1=1$ and $\varepsilon_2=-44.05+3.24i$, closely match the experimental data. Clearly, the dipolar model is again fully consistent with the near-field structure depicted in real-world experiments.

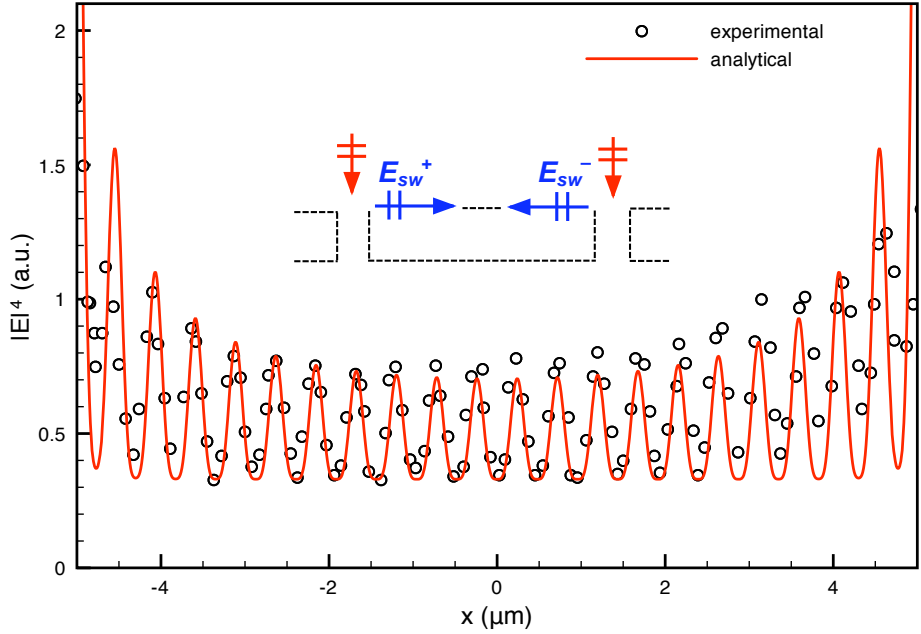

Fig. 7. Near-field fringe pattern in the double-slit setup. Comparison between the experimental (circles) [7] and analytical (solid line) results with $2d=10.44\mu m$, $\lambda=974.3nm$ and $\varepsilon_2 = -44.05+3.24i$

It is worth to emphasize that the two studies [3,7] reported in this Section both assess the SW field in the near-zone propagation regime ($0<x<2L_{bw}$). The corresponding results yielded for either investigation by the analytical dipolar model indicate that the asymptotic solution is fairly accurate inside this transient near-zone as expected from the accuracy range $k_0x>2$, or stated alternatively, $x>\lambda/\pi$ (see Subsection 3.1).

5.  Conclusion

In summary, we provide a rigorous closed-form description of the surface wave generated via the diffraction of a horizontal Hertzian dipole over the air-metal interface by solving the corresponding Sommerfeld integral with a modified method of steepest descents. The asymptotic solution – accurate for distances $x>\lambda/\pi$ from the origin – demonstrates that the total surface near field is composed of two distinct components as previously evidenced in [6]: a long-range surface plasmon polariton (SPP) evanescent wave and a short-range boundary wave (BW) with free-space cylindrical properties. The dynamics of both constituent waves are fully revealed through simple analytic expressions, and appropriate parameters are defined to quantitatively and continuously describe their properties across the entire propagation range: near and asymptotic. Moreover, our calculations predict that the wavevector mismatch of the two co-propagating surface modes creates a weak periodic beating of the total field amplitude, which is noticeable at relatively large distances from the source and high optical frequencies. The closed-form solution is further validated via comparison between a dipolar model and recent experimental data, which demonstrates excellent quantitative agreement. In the process we show that the generation of surface waves by diffraction – and their ensuing interactions between one-dimensional subwavelength-sized nano-defects along the metallo-dielectric interface – are conveniently and accurately modeled with linear Hertzian dipoles. Hence, the theoretical formalism presented in this contribution can be used for the characterization and engineering of the near-field interactions in plasmonic and nano-optical devices while alleviating the reliance on time-consuming computer simulations.

**Acknowledgments**

The authors cordially thank Robert E. Collin from Case Western Reserve University for his valuable remarks and suggestions. Helpful discussions with J. Weiner of IRSAMC/LCAR and P. Lalanne of the Institut d'Optique are also gratefully acknowledged.

______________________________________________